\documentclass[conference]{IEEEtran}
\IEEEoverridecommandlockouts

\usepackage{cite}
\usepackage{amsmath,amssymb,amsfonts,amsthm}
\usepackage{algorithmic}
\usepackage{graphicx}
\usepackage{textcomp}
\usepackage{xcolor}
\usepackage{balance}
\usepackage{booktabs}
\usepackage{url}
\usepackage[hidelinks]{hyperref}
\usepackage{algorithmic}
\usepackage{graphicx}
\usepackage{placeins}
\usepackage{textcomp}
\usepackage{xcolor}
\usepackage{tikz}
\usetikzlibrary{arrows.meta,positioning,calc,decorations.pathreplacing}
\tikzset{
	vtx/.style   = {circle,draw,minimum size=5.2mm,inner sep=0pt,font=\small},
	slot/.style  = {draw,rounded corners=1pt,minimum width=5.4mm,minimum height=5.0mm,
		inner sep=0pt,font=\scriptsize},
	hslot/.style = {slot,fill=black!12},
	sent/.style  = {slot,fill=black!25,text=black!70},
	dead/.style  = {slot,text=black!35,draw=black!30,densely dotted},
	arr/.style   = {-{Latex[length=1.5mm]},thin},
	gth/.style   = {arr,densely dashed,black!65},
	lbl/.style   = {font=\scriptsize},
	tny/.style   = {font=\tiny},
	brc/.style   = {decorate,decoration={brace,amplitude=3pt,raise=1pt,mirror},black!55},
}

\newcommand{\TGraph}{\mathcal{G}}
\newcommand{\TE}{\mathcal{E}}
\newcommand{\Hop}{\mathcal{H}}
\newcommand{\Nout}{\mathcal{N}^{+}}
\newcommand{\hin}[3]{h^{(#3)}_{#1,#2,+}}
\newcommand{\hout}[2]{h^{(#2)}_{#1,+}}
\newcommand{\set}[1]{\left\{#1\right\}}
\newcommand{\mset}[1]{\left\{\!\!\!\left\{#1\right\}\!\!\!\right\}}

\newcommand{\Dout}{\Delta^{+}_{\max}}

\newcommand{\Warp}{\omega}
\newcommand{\Blk}{B}

\newcommand{\cnt}{\mathit{cnt}}
\newtheorem{theorem}{Theorem}
\newtheorem{lemma}[theorem]{Lemma}
\newtheorem{proposition}[theorem]{Proposition}

\newtheorem{example}[theorem]{Example}

\newcommand{\CpuHundredFourSpeedup}{69.1\ensuremath{\times}}

\newcommand{\LargeMultiSpeedupRange}{6.21--7.52\ensuremath{\times}}

\newcommand{\LargeMultiBillionTransferShare}{26.8\%}

\newcommand{\RedditOrderOneTwentyEightEightVsCpuMt}{7.85\ensuremath{\times}}
\newcommand{\CompressionComputeRange}{1.01--1.10\ensuremath{\times}}

\newcommand{\CompressionStaticSavingRange}{15.5--30.4\%}
\newcommand{\CompressionWorkingSavingRange}{32.2--63.2\%}

\newcommand{\RedditOptimizedVsCpuMtEte}{1.48\ensuremath{\times}}

\newcommand{\FrontierRedditActiveAtOneTwentyEight}{95.6\%}

\newcommand{\RedditStorageFullOverCanonical}{0.983\ensuremath{\times}}
\newcommand{\RedditStorageFullOverCanonicalCi}{0.960--1.054\ensuremath{\times}}

\newcommand{\RedditOptimizedEteSpeedup}{1.07\ensuremath{\times}}
\newcommand{\RedditOptimizedEteCi}{1.02--1.11\ensuremath{\times}}

\newcommand{\RedditProcessCi}{0.97--1.06\ensuremath{\times}}

\newcommand{\ExchangeTimeReddit}{1.035\ensuremath{\times}}
\newcommand{\ExchangeTimeBillion}{1.099\ensuremath{\times}}

\newcommand{\ExchangeSavedReddit}{72.0\%}
\newcommand{\ExchangeSavedBillion}{81.2\%}
\newcommand{\KernelNoWarpSlowdownRange}{1.01--1.89\ensuremath{\times}}
\newcommand{\KernelNoEarlySlowdownRange}{0.994--1.014\ensuremath{\times}}

\newcommand{\TransitionRoundRatioRange}{0.994--1.000\ensuremath{\times}}

\begin{document}

\title{Parallel Algorithms for Higher-Order Temporal H-Index Computation}
\author{\IEEEauthorblockN{Lutz Oettershagen}
\IEEEauthorblockA{\textit{School of Computer Science and Informatics} \\
\textit{University of Liverpool}\\
Liverpool, UK \\
\url{lutz.oettershagen@liverpool.ac.uk}}
}

\maketitle

\begin{abstract}
	Ranking vertices by importance is a fundamental task in network mining. The H-index provides such a ranking by asking how many neighbors are themselves well connected; its higher-order temporal variant recursively applies this idea over time-respecting neighborhoods in temporal networks.
	Scaling its computation is difficult because the existing streaming algorithm intertwines temporal processing with H-index order, incurring maximum-degree-dependent costs and exposing little parallelism. We show that the temporal dependencies can instead be resolved once: each edge's time-dependent lookup is mapped to a fixed successor position, turning every H-index order into a parallel gather followed by a segmented suffix-H computation. Based on this reformulation, we derive a linear-work suffix-H algorithm, a parallel multicore CPU implementation, degree-adaptive GPU kernels, and a multi-GPU scheme that exchanges only distinct remotely queried states. Experiments on synthetic and real-world temporal graphs show that the reformulation alone is substantially faster than the streaming algorithm, and that GPUs add order-round speedups exceeding an order of magnitude over our multicore implementation on synthetic inputs with up to one billion temporal edges. On a 636M-edge Reddit network, eight GPUs compute 128 orders in 3.6 s of round time, whereas the streaming algorithm does not finish within an hour even at order 8.\looseness=-1
\end{abstract}


\section{Introduction}
\label{sec:introduction}

Identifying and ranking important vertices is a fundamental task in network mining and graph analytics, with applications ranging from influence analysis to the identification of key actors in spreading processes~\cite{FREEMAN1978215,LU20161}. In temporal networks, however, vertex importance depends not only on who is connected, but also on when interactions occur. Interactions such as messages, contacts, and transactions happen at specific times, and temporal graphs capture this information so that paths and information flows respect the order in which events occur~\cite{holme2015modern,Michail16,streamgraphs}.\looseness=-1

The higher-order temporal H-index~\cite{oettershagen2023higherorder} measures vertex importance under such time-respecting reachability. It extends the classical H-index and its higher-order variants~\cite{hirsch2005index,lu2016h} by recursively evaluating the importance of temporally reachable neighbors. This makes it particularly relevant for identifying influential vertices in spreading processes: it has been shown to be an effective heuristic for detecting potential super-spreaders in evolving social networks and has subsequently been considered in the broader study of centrality and influence in temporal networks~\cite{oettershagen2025edge,andriamampianina2025selective,mao2026historical}.

Computing this index at scale is challenging because every temporal edge
induces a time-dependent lookup into the state of another vertex. The existing
 streaming algorithm (\textsc{Stream}) proposed in~\cite{oettershagen2023higherorder} exploits a reverse chronological edge order, but
interleaves time and H-index order in a single streaming computation. This
organization introduces maximum-degree-dependent work and state and exposes
little coarse-grained parallelism. It also assumes uniform edge transition
times. These limitations become increasingly important for large temporal
graphs and modern parallel hardware.

Our key observation is that the recurrence has a simpler data-access structure
when organized by H-index order rather than by time. At order $i$, every
temporal edge queries only a state from order $i-1$, evaluated at the arrival
time of its head vertex. Once these temporal access locations are materialized,
an order consists of two bulk operations: a gather over temporal edges and,
independently for each vertex, an H-index computation for every suffix of its
time-sorted adjacency list. We capture the static access pattern with a
temporal CSR representation and successor indices, while keeping the
order-dependent state separate. 
Our central contribution is a reformulation of the temporal recurrence that improves sequential complexity, exposes parallelism across edges and vertices, and identifies the states needed between orders. Resolving temporal access locations once yields a fixed gather–suffix dataflow; computing all suffix H-indices in linear work removes the maximum-degree factor from sequential round work. The same static query structure determines which states must persist and be exchanged across partitions, providing the basis for our CPU, GPU, and multi-GPU algorithms.

The remaining challenge is to evaluate all suffix H-indices efficiently.
Naively treating the suffixes independently repeats substantial work. We show
instead that all suffix-H values of a segment can be computed in linear
sequential work, yielding an $O(n|\TE|)$ algorithm, where $n$ is the order and $\TE$ the set of temporal edges, after preprocessing.
For GPUs, temporal degree skew requires further care: most adjacency segments
are short, whereas a small number of hubs may contain thousands of temporal
edges. We therefore use degree-adaptive kernels that combine warp-level
processing for short segments with tiled block processing for larger ones.
The same formulation extends to arbitrary positive transition times and to
multiple GPUs, where only the canonical states actually queried across
partitions need to be communicated.

\textbf{Our contributions are as follows:}

(i) We derive an order-major formulation of the higher-order temporal
H-index. Temporal lookups are resolved once into static successor indices,
so each order becomes a parallel gather followed by a segmented suffix-H
computation; nonuniform positive transition times are handled entirely in
the successor construction.\looseness=-1

(ii) We adapt a single-insertion histogram technique for H-index
maintenance to the setting where every suffix of a segment is required,
giving a linear-work segmented suffix-H algorithm. This yields an
$O(n|\TE|)$ sequential algorithm.

(iii) We design a new parallel multicore CPU implementation and
degree-adaptive GPU algorithms, both with $O(|V|+|\TE|)$ working memory.
The GPU algorithms include warp- and block-level CUDA kernels for highly
skewed temporal degree distributions.

(iv) We extend the method to multiple GPUs using an owner-computes
decomposition with deduplicated, query-aware halos and full or change-only
direct state exchange.

(v) We evaluate order-round and full-pipeline performance on real graphs and
on controlled graphs with up to one billion temporal edges, and identify
the graph sizes and degree distributions for which each platform is
preferable.

\section{Related Work}
\label{sec:related}
Temporal-network algorithms and centrality measures are surveyed in
~\cite{holme2015modern,streamgraphs,santoro2011time,Michail16}. Prior work
includes temporal closeness and betweenness
~\cite{pan2011path,magnien2015time,oettershagen2020efficient,crescenziMM20topk,BussMNR20,santoro2022onbra,tsalouchidou2019temporal},
walk-based centralities
~\cite{katztg,grindrod2011communicability,RozenshteinG16,oettershagen2022temporal},
and temporal core notions
~\cite{wu2015core,galimberti2020span,hung2021maximum,qin2019mining,qin2020mining,li2018persistent}.
The temporal H-index differs by recursively aggregating strong time-respecting
neighbors into one value per vertex and order.
The original \textsc{Stream} algorithm processes temporal edges in reverse
chronological order and computes all indices up to order $n$ in
$O(mn\Dout)$ time and $O(|V|n\Dout)$ space
~\cite{oettershagen2023higherorder}. Repeated H-index iteration on static graphs
is closely related to core and nucleus decompositions
~\cite{lu2016h,sariyuce2018local}. The at-most-one H-index change under one
insertion and incremental histograms are standard; we do not claim them as
new. Distributed and multicore $k$-core methods~\cite{montresor2013distributed,
kabir2017parallel} instead parallelize static peeling. Our temporal workload
requires an arrival-time-dependent lower-order state for every edge; the
static-successor reformulation exposes these queries for parallel execution.
Degree-aware thread/warp/block assignment is standard in GPU graph processing
~\cite{wang2017gunrock}; GPU $k$-core implementations use parallel peeling
and frontiers~\cite{tripathy2018scalable,ahmad2023gpu}. Other work addresses
dynamic storage, memory pressure, distributed operators, and host--GPU movement
~\cite{sha2017dynamicgpu,gera2020uvm,fan2021graphscope,wang2023hytgraph}.
We use degree tiers for temporal skew, but the specific contributions are
static successors, breakpoint-based suffix kernels, and query-aware halos.\looseness=-1

\section{Preliminaries}
\label{sec:preliminaries}

A directed temporal network is $\TGraph=(V,\TE)$, where a temporal edge
$e=(u,w,t_e,\lambda_e)$ is available at time $t_e\ge0$ and reaches $w$ at
$a_e=t_e+\lambda_e$, with $\lambda_e>0$. For vertex $v$ and query time $t$,
\begin{equation}
    \Nout(v,t)=
    \mset{(w,a_e):e=(v,w,t_e,\lambda_e)\in\TE,\ t_e\ge t},
    \label{eq:neighborhood}
\end{equation}
with temporal out-degree $\delta^+(v,t)=|\Nout(v,t)|$. Distinct temporal edges
contribute with multiplicity. We write $m=|\TE|$, let $d_v^+$ be the number of
outgoing temporal edges of $v$, and set $\Dout=\max_v d_v^+$.

\textbf{Higher-order temporal H-index.}
For a finite multiset $S$ of nonnegative integers,
\begin{equation}
    \Hop(S)=\max\set{k\in\mathbb{N}_0:|\set{x\in S:x\ge k}|\ge k},
    \label{eq:hop}
\end{equation}
with $\Hop(\emptyset)=0$. 
The outward temporal H-index of vertex $v$ at query time $t$ is defined for $i>0$ 
recursively by
\begin{align}
	\hin{v}{t}{i}
	\!\!=\!
	\Hop\!\left(\!\!
	\mset{\hin{w}{a_e}{i-1}:
		e=(v,w,t_e,\lambda_e)\in\TE,\ t_e\ge t}
	\!\!\right)\!,\label{eq:recurrence}
\end{align}
with base case $\hin{v}{t}{0}=\delta^+(v,t)$.
The order-$i$ score of $v$ is $\hout{v}{i}=\hin{v}{0}{i}$. Order 0 is
the temporal out-degree, while order one is the temporal analogue of the
classical graph H-index.

\textbf{Problem:} Given a temporal network $\TGraph$ and an order $n$, our computational problem
is to compute $\hout{v}{n}$ for every $v\in V$.\footnote{
\textbf{Inward variant:} The inward index of~\cite{oettershagen2023higherorder}
does not require a second algorithm. Let
$T=\max_{e\in\TE}(t_e+\lambda_e)$ and replace every edge
$(u,v,t,\lambda)$ by $(v,u,T-(t+\lambda),\lambda)$. The reversed edge arrives
at $T-t$, so an original edge that can reach $v$ by time $t'$ corresponds
exactly to a reversed edge leaving $v$ no earlier than $T-t'$. Consequently the
inward recurrence on $\TGraph$ is the outward recurrence on this
transition-aware transpose after the time change $t'\mapsto T-t'$. We
therefore focus on the outward case.}
The $t_e\ge t$ availability condition, treatment of tied temporal events,
and inward reduction match~\cite{oettershagen2023higherorder}.

\section{Order-Major Reformulation}
\label{sec:reformulation}

The recurrence has two sources of repeated work: each temporal edge must locate
a lower-order state at its arrival time, and each vertex requires H-indices for
many overlapping suffixes. We first materialize all arrival-time accesses once,
which exposes an order-major gather--suffix recurrence; we then compute all
suffix H-indices in linear sequential work and identify the states that must
persist between orders.

\subsection{Static Temporal Access Structure}
\label{sec:temporal-access}

For the recurrence~\eqref{eq:recurrence}, traversing an edge
$e=(u,w,t_e,\lambda_e)$ requires the lower-order state of $w$ at arrival time
$a_e=t_e+\lambda_e$. A direct implementation could locate this state by
searching the temporal adjacency list of $w$ at every order. Since the queried
arrival times depend only on the input graph, however, all such locations can
be materialized once beforehand.

For brevity, let $d_v=d_v^+$ denote the number of outgoing temporal edges of
$v$. We sort these edges by nondecreasing availability time, breaking
equal-time ties arbitrarily, and denote the edge at local position $p$ by
$e_{v,p}$, where $0\le p<d_v$. The segments are concatenated in a temporal
compressed-sparse-row (TCSR) representation with offset array
$\mathit{off}[0\ldots |V|]$.

Every temporal neighborhood query is a suffix of one such segment. Let
\[
p_v(t)=
\operatorname{lower\_bound}
(t_{e_{v,0}},\ldots,t_{e_{v,d_v-1}},t).
\]
where $\operatorname{lower\_bound}$ returns the smallest position $p$ such that
$t_{e_{v,p}}\ge t$, or $d_v$ if no such position exists.
Then $\Nout(v,t)$ consists exactly of the edges at positions
$p_v(t),\ldots,d_v-1$.
Consequently, for an edge
$e=(u,w,t_e,\lambda_e)$ we precompute the \emph{successor index}
\begin{equation}
	\mathit{succ}[e]=p_w(a_e)=p_w(t_e+\lambda_e).
	\label{eq:succ}
\end{equation}
We use the convention that a lookup at position $d_w$ returns 0.

The successor index is static: it depends on the temporal graph but not on the
H-index order. Thus, after preprocessing, following a temporal edge requires
only an array lookup rather than a temporal search. This separation between
static temporal access and order-dependent state is the basis of the
order-major formulation below.

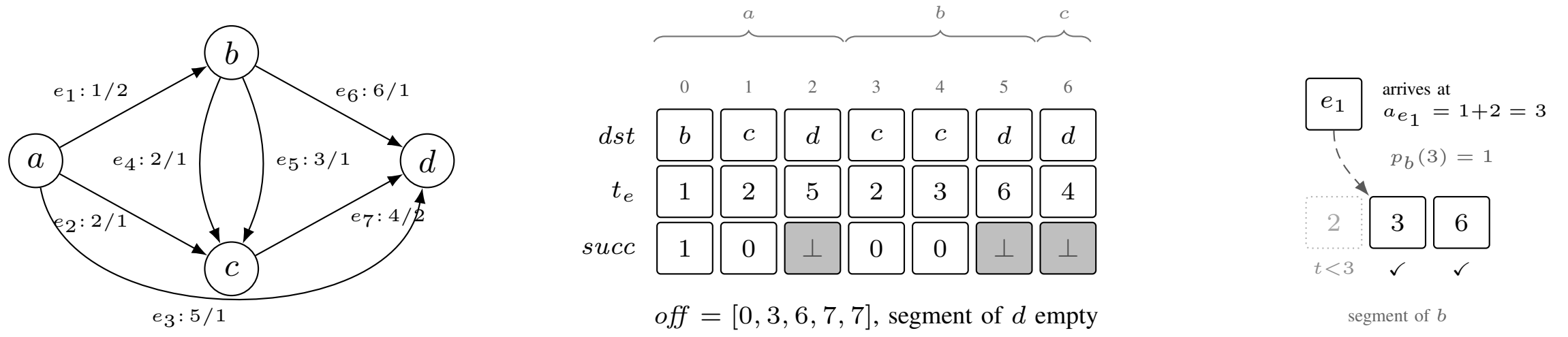
\begin{figure*}[t]
	\centering
	\begin{tikzpicture}
		\begin{scope}[local bounding box=PA]
			\node[vtx] (a) at (0,0)      {$a$};
			\node[vtx] (b) at (1.9,1.05) {$b$};
			\node[vtx] (c) at (1.9,-1.05){$c$};
			\node[vtx] (d) at (3.8,0)    {$d$};
			\draw[arr] (a) -- node[tny,above left=-3pt] {$e_1{:}\,1/2$} (b);
			\draw[arr] (a) -- node[tny,left=2pt,pos=.60] {$e_2{:}\,2/1$} (c);
			\draw[arr] (a) to[bend right=80]
			node[tny,below,pos=.42] {$e_3{:}\,5/1$} (d);
			\draw[arr] (b) to[bend right=24]
			node[tny,left=0pt,pos=.5] {$e_4{:}\,2/1$} (c);
			\draw[arr] (b) to[bend left=24]
			node[tny,right=0pt,pos=.5] {$e_5{:}\,3/1$} (c);
			\draw[arr] (b) --
			node[tny,above right=-3pt] {$e_6{:}\,6/1$} (d);
			\draw[arr] (c) --
			node[tny,below right=-4pt,pos=.62] {$e_7{:}\,4/2$} (d);
		\end{scope}
		\node[lbl,below=0pt of PA]
		{(a) temporal network, edge labels $e{:}\,t_e/\lambda_e$};

		\begin{scope}[shift={(6.3,-.60)},local bounding box=PB]
			\draw[brc,yshift=1.72cm] (1.54,0) -- (-0.30,0)
			node[midway,above=4pt,tny] {$a$};
			\draw[brc,yshift=1.72cm] (3.40,0) -- (1.56,0)
			node[midway,above=4pt,tny] {$b$};
			\draw[brc,yshift=1.72cm] (3.94,0) -- (3.42,0)
			node[midway,above=4pt,tny] {$c$};
			
			\foreach \p in {0,...,6}{
				\node[tny,black!60] at (\p*0.62,1.32) {\p};
			}
			\foreach \p/\v in {0/b,1/c,2/d,3/c,4/c,5/d,6/d}{
				\node[slot] (D\p) at (\p*0.62,0.85) {$\v$};
			}
			\foreach \p/\v in {0/1,1/2,2/5,3/2,4/3,5/6,6/4}{
				\node[slot] (T\p) at (\p*0.62,0.30) {$\v$};
			}
			\foreach \p/\v in {0/1,1/0,3/0,4/0}{
				\node[slot] (S\p) at (\p*0.62,-0.25) {$\v$};
			}
			\foreach \p in {2,5,6}{
				\node[sent] (S\p) at (\p*0.62,-0.25) {$\bot$};
			}
			
			\node[lbl,left=2pt of D0] {$\mathit{dst}$};
			\node[lbl,left=2pt of T0] {$t_e$};
			\node[lbl,left=2pt of S0] {$\mathit{succ}$};
			\node[lbl] at (1.85,-0.92)
			{$\mathit{off}=[0,3,6,7,7]$, segment of $d$ empty};
		\end{scope}
		\node[lbl,below=0pt of PB] {(b) temporal CSR};

		\begin{scope}[shift={(12.6,-0.60)},local bounding box=PC]
			\node[slot] (e1) at (0,1.15) {$e_1$};
			\node[tny,right=2pt of e1,align=left]
			{arrives at\\$a_{e_1}=1{+}2=3$};
			
			\foreach \p/\v in {1/3,2/6}{
				\node[slot] (B\p) at (\p*0.62,0) {$\v$};
			}
			\node[dead] (B0) at (0,0) {$2$};
			\node[tny,black!45] at (0,-0.45) {$t{<}3$};
			\node[tny] at (0.62,-0.45) {$\checkmark$};
			\node[tny] at (1.24,-0.45) {$\checkmark$};
			\node[tny,black!60] at (0.62,-0.92) {segment of $b$};
			\draw[gth] (e1.south) to[out=-90,in=130] (B1.north west);
			\node[tny,black!65] at (1.05,0.62) {$p_b(3)=1$};
		\end{scope}
		\node[lbl,below=0pt of PC] {(c) successor index of $e_1$};
	\end{tikzpicture}%
	\caption{Running temporal network.
		Shaded $\bot$ entries denote zero sentinels; here they arise from
		the empty outgoing segment of $d$. The dotted slot in~(c) is excluded
		because its time precedes the arrival time of $e_1$, while the two later
		slots remain eligible.}
	\label{fig:running}
\end{figure*}

\subsection{Order-Major Recurrence}
\label{sec:order-major-recurrence}

Once successor positions are fixed, the temporal recursion can be reorganized
by H-index order. The key observation is that every value at order $i$ depends
only on values from order $i-1$; there are no dependencies among values
produced within the same order. We therefore materialize the state of one
order, use it to evaluate all temporal-edge contributions to the next order,
and then compute the required suffix H-indices independently for each vertex.

For every vertex $v$, position $0\le p<d_v$, and order $i$, let $F_i[v,p]$
denote the H-index state associated with the suffix beginning at position $p$,
and set $F_i[v,d_v]=0$. Define
\begin{align}
	F_0[v,p]
	&=d_v-p,
	\label{eq:state-base}\\
	z_i[e_{v,j}]
	&=
	F_{i-1}[w,\mathit{succ}[e_{v,j}]],
	\qquad
	e_{v,j}=(v,w,t,\lambda),
	\label{eq:state-gather}\\
	F_i[v,p]
	&=
	\Hop\left(
	\mset{z_i[e_{v,j}]:p\le j<d_v}
	\right),
	\qquad i\ge1.
	\label{eq:state-suffix}
\end{align}

The array $F_i[v,*]$ is defined at every position because doing so yields a
simple contiguous representation and permits suffix processing. Only
positions of the form $p_v(t)$, however, correspond directly to temporal
queries. This distinction matters when several edges have the same
availability time: $p_v(t)$ always selects the first position of the tied
group.

\begin{theorem}
	\label{thm:canonical}
	For every temporal network with positive, possibly nonuniform transition
	times, every vertex $v$, query time $t$, and order $i\ge0$,
	$
	F_i[v,p_v(t)]=\hin{v}{t}{i},
	$
	where $F_i[v,d_v]=0$ by definition. Consequently,
	$\hout{v}{i}=F_i[v,0]$ for every vertex $v$, including $d_v=0$, where both
	sides vanish.
\end{theorem}

\begin{proof}
Let $p=p_v(t)$. By construction, positions $p,\ldots,d_v-1$ are exactly the
edges available at or after $t$, so the claim holds for $i=0$. For $i\ge1$,
\eqref{eq:succ} and the induction hypothesis give
$z_i[e]=\hin{w}{t_e+\lambda_e}{i-1}$ for every edge in that suffix. Hence, applying
$\Hop$ gives~\eqref{eq:recurrence}. Lower bounds select the
first edge of an equal-time group, and $p_v(0)=0$, giving
$\hout{v}{i}=F_i[v,0]$.
\end{proof}

Theorem~\ref{thm:canonical} gives two independent phases per order: each edge
gathers one previous-order state through its successor index, then each vertex
computes suffix-H on its own segment. No order-$i$ value is read within that
order, so only consecutive orders need synchronization.

\begin{lemma}%
	\label{lem:order-monotonicity}
	For every vertex $v$, suffix position $0\le p\le d_v$, and order $i\ge1$,
	$F_i[v,p]\le F_{i-1}[v,p]$.
\end{lemma}

\begin{proof}
	For $i=1$, $F_1[v,p]$ is the H-index of a multiset containing
	$d_v-p=F_0[v,p]$ values and therefore cannot exceed $F_0[v,p]$.
	Assume the claim holds at order $i-1$. Every gathered contribution at
	order $i$ is then at most the corresponding contribution at order $i-1$.
	The H-index operator is monotone under componentwise decreases, so applying
	it to any suffix gives $F_i[v,p]\le F_{i-1}[v,p]$. The sentinel position is
	zero at every order.
\end{proof}

\subsection{Linear-Work Suffix-H}
\label{sec:linear-suffix}

The order-major recurrence removes repeated temporal searches, but leaves one
nonstandard primitive: for every vertex, we must compute an H-index for every
suffix of its gathered values. Computing these suffixes independently repeats
work because adjacent suffixes differ by only one element. We exploit this
single-insertion structure to compute all suffix values in linear work.

For an array $z_0,\ldots,z_{d-1}$, define
$
r_p=\Hop(\mset{z_j:p\le j<d}).
$
When the suffix is extended from right to left, exactly one value is inserted.
The following property shows that its H-index can increase by at most one.

\begin{lemma}
	\label{lem:insertion}
	Let $S$ be a finite multiset, let $h=\Hop(S)$, and let
	$S'=S\cup\mset{x}$. Then
	$
	\Hop(S')\in\{h,h+1\}.
	$
	Moreover,
	\[
	\Hop(S')=h+1
	\quad\Longleftrightarrow\quad
	|\set{y\in S':y\ge h+1}|\ge h+1.
	\]
\end{lemma}

\begin{proof}
	Adding an element cannot decrease the H-index, so $\Hop(S')\ge h$.
	Suppose for contradiction that $\Hop(S')\ge h+2$. Then $S'$ contains at
	least $h+2$ elements of value at least $h+2$. Removing the newly inserted
	element leaves at least $h+1$ such elements in $S$, implying
	$\Hop(S)\ge h+1$, a contradiction. Hence $\Hop(S')\le h+1$.
	
	The only possible increase is therefore from $h$ to $h+1$. By definition
	of $\Hop$, this increase occurs exactly when $S'$ contains at least $h+1$
	elements whose values are at least $h+1$.
\end{proof}

Lemma~\ref{lem:insertion} allows a right-to-left sweep to maintain only the
current H-index together with enough information to test its next possible
threshold. Figure~\ref{fig:sequential-suffix} gives the resulting procedure.

\begin{figure}[t]
	\small
	\begin{algorithmic}[1]
		\REQUIRE Nonnegative integers $z_0,\ldots,z_{d-1}$
		\ENSURE $r_p=\Hop(\mset{z_j:j\ge p})$ for every $p$
		\STATE $\mathit{freq}[0\ldots d]\gets0$;
		$h\gets0$;
		$\cnt\gets0$
		\FOR{$p\gets d-1$ downto $0$}
		\STATE $x\gets\min\{z_p,d\}$;
		$\mathit{freq}[x]\gets\mathit{freq}[x]+1$
		\IF{$x\ge h+1$}
		\STATE $\cnt\gets\cnt+1$
		\ENDIF
		\IF{$\cnt\ge h+1$}
		\STATE $h\gets h+1$
		\STATE $\cnt\gets\cnt-\mathit{freq}[h]$
		\ENDIF
		\STATE $r_p\gets h$
		\ENDFOR
		\STATE \textbf{return} $r$
	\end{algorithmic}
	\caption{\textsc{SequentialSuffixH}. The variable $\cnt$ counts values at
		least $h+1$ in the currently processed suffix.}
	\label{fig:sequential-suffix}
\end{figure}

\begin{theorem}
	\label{thm:linear-suffix}
	The procedure in Figure~\ref{fig:sequential-suffix} computes all $d$ suffix-H values in
	$\Theta(d)$ time and $O(d)$ auxiliary space.
\end{theorem}

\begin{proof}
	Clamping at $d$ is safe because no suffix H-index exceeds $d$. After each
	insertion, $\mathit{freq}$ counts current-suffix values, $h$ is its
	H-index, and $\cnt$ counts values at least $h+1$. The histogram and
	$\cnt$ update for $z_p$; Lemma~\ref{lem:insertion} permits only an
	$h\to h+1$ change, exactly when $\cnt\ge h+1$. Subtracting
	$\mathit{freq}[h]$ after that change restores the count above the next
	threshold. Each iteration is constant time after $O(d)$ initialization;
	writing $d$ results gives the matching lower bound, and the histogram
	uses $O(d)$ space.
\end{proof}

Combining the static successor accesses with the linear suffix algorithm gives
the overall sequential complexity of the order-major formulation.

\begin{theorem}
	\label{thm:sequential-order-major}
	After the TCSR representation and successor indices have been constructed,
	the outward temporal H-index of order $n$ for all vertices can be computed
	in $O(n(m+|V|))$ time and $O(m+|V|)$ space, where $m=|\TE|$. If isolated
	vertices are omitted, the time bound is $O(nm)$. The algorithm supports
	arbitrary positive transition times.
\end{theorem}

\begin{proof}
	Initialize $F_0$ according to~\eqref{eq:state-base}. At each order,
	equation~\eqref{eq:state-gather} performs one constant-time lookup per
	temporal edge, for $O(m)$ work. Applying \textsc{SequentialSuffixH} (Figure~\ref{fig:sequential-suffix})
	to the segment of each vertex $v$
	requires $O(d_v)$ time; summing over all vertices gives
	$
	\sum_{v\in V} O(d_v)=O(m).
	$
	Visiting the vertex descriptors contributes $O(|V|)$ per order.
	Theorem~\ref{thm:canonical} establishes correctness of the resulting
	states.
	
	Only the previous and current orders need to coexist. Two $m$-entry state
	arrays suffice, together with another $m$ entries if the gathered values are
	materialized explicitly. Including the TCSR metadata gives
	$O(m+|V|)$ space. If isolated vertices are omitted, then $|V|\le2m$ and
	the time bound simplifies to $O(nm)$. Nonuniform transition times affect
	only the precomputed successor positions and therefore do not change the
	per-order bound.
\end{proof}

With comparison sorting and binary search for successor positions,
preprocessing takes $O(m\log m)$ time: each successor search costs at most
$O(\log(\Dout+1))=O(\log m)$ and there are $m$ searches. Thus, the complete
computation takes
\[O\left(m\log m+n(m+|V|)\right)\]
time and $O(m+|V|)$ space.

\subsection{Canonical States}
\label{sec:canonical-states}

The recurrence above conceptually defines $F_i[v,p]$ at every TCSR position.
Most of these values, however, are never read after the current order. Across
an order boundary, a state is needed only if it is either the position-zero
score of a vertex or the target of at least one successor query. We call these
positions \emph{canonical states}.
For each vertex $v$, let
\[
P_v=
\set{0}
\cup
\set{
	\mathit{succ}[e]:
	e\in\TE,\ 
	\operatorname{head}(e)=v,\ 
	\mathit{succ}[e]<d_v
},
\]
and let
$
Q=\sum_{v\in V}|P_v|.
$
Repeated successor positions appear only once in $P_v$.

\begin{proposition}
	\label{prop:compression}
	It is sufficient to retain $F_i[v,p]$ between orders only for
	$p\in P_v$. Thus, two persistent order arrays require $O(Q)$ entries,
	where $Q\le m+|V|$. A round still uses $O(m+Q)$ work. Canonical
	compression reduces persistent state, not the edge-sized transient gather:
	if the gathered values are materialized, the round uses $\Theta(m)$
	entries for that buffer in addition to the $O(Q)$ persistent state.
\end{proposition}

\begin{proof}
	Position zero is required for the reported vertex score. Every other
	cross-order read has the form
	$
	F_{i-1}[w,\mathit{succ}[e]],
	$
	and therefore accesses a position in $P_w$, unless
	$\mathit{succ}[e]=d_w$, in which case the value is the implicit zero
	sentinel. No other state is queried across an order boundary.
	
	During a round, each complete segment is still scanned from right to left,
	as in \textsc{SequentialSuffixH} (Figure~\ref{fig:sequential-suffix}), but a suffix value need only
	be retained when its position belongs to $P_v$. Since the scan incorporates
	every edge contribution, all retained values remain exact. Induction over
	the orders therefore establishes sufficiency.
	
	There is at most one successor position contributed by each temporal edge,
	plus position zero for each vertex, so $Q\le m+|V|$. The edge gather and
	suffix scans require $O(m)$ work, while initializing or writing retained
	states requires $O(Q)$ work.
\end{proof}

Canonical states separate transient suffix computation from persistent
order-to-order state. This distinction can substantially reduce persistent
storage when many temporal edges share the same successor position. More
importantly, it identifies exactly which states may be queried across a
partition boundary in the multi-GPU algorithm developed later: communication
depends on distinct canonical queries rather than directly on the number of
cross-partition temporal edges.

\section{Order-Major GPU Algorithm}
\label{sec:algorithm}

Section~\ref{sec:reformulation} reduces each H-index order to two bulk
operations: a gather over temporal edges and a suffix-H computation within
each vertex segment. The gather is naturally data parallel: every temporal
edge performs one independent lookup into the previous-order state. The
suffix-H phase is more challenging. The \textsc{SequentialSuffixH} procedure
in Figure~\ref{fig:sequential-suffix} achieves linear sequential work by scanning each segment from right to left,
but its H-index state is loop-carried and therefore does not directly expose
parallelism. Moreover, temporal degree distributions can be highly skewed, so
a single GPU execution strategy is inefficient across all segment sizes.

Our GPU implementation addresses these issues with degree-adaptive segmented
suffix-H kernels. Short segments are processed entirely within a warp using
ballots and population counts, while longer segments are processed by blocks
in right-to-left tiles. Both kernels use a breakpoint representation that lets
many suffixes share the work of testing H-index thresholds. Vertices are
partitioned into degree-tier worklists so that each segment uses an
appropriate execution and memory strategy.

\subsection{Execution Overview}
\label{sec:gpu-overview}

Preprocessing constructs the TCSR, successor indices, canonical-state map, and
degree-tier worklists once. During order $i$, a grid-stride gather assigns one
thread per temporal edge and evaluates
\[
x[e]=F_{\mathrm{cur}}[w,\mathit{succ}[e]].
\]
The gathered values retain the TCSR segmentation; each nonempty segment is then
processed by the appropriate suffix-H kernel, canonical positions are written
to $F_{\mathrm{next}}$, and the state buffers are swapped. Sentinel successors
are implicit zeros. Because every temporal lookup is fixed before the first
round, nonuniform positive transition times introduce no control flow or search
inside the order rounds.

\subsection{Parallel Suffix-H}
\label{sec:gpu-suffix}

The linear-work algorithm of Section~\ref{sec:linear-suffix} maintains the
H-index of the processed suffix while inserting one value at a time from
right to left. This dependency is ideal sequentially but prevents different
positions of a segment from being evaluated independently. For GPU execution,
we therefore use an equivalent representation based on the positions at which
H-index thresholds become feasible.
For one vertex, let $z_0,\ldots,z_{d-1}$ be its gathered values and define
\[
r_p=\Hop(\mset{z_j:j\ge p}).
\]
For each threshold $k\ge1$, define its \emph{breakpoint}
\begin{equation}
	b_k=
	\max\set{
		p:
		|\set{j\ge p:z_j\ge k}|\ge k
	},
	\label{eq:breakpoint}
\end{equation}
with $b_k=-1$ if no suffix reaches threshold $k$.

\begin{lemma}
	\label{lem:breakpoint}
	The sequence $b_1,b_2,\ldots$ is non-increasing, and for every position $p$,
		$r_p=
		|\set{k\ge1:b_k\ge p}|.$
\end{lemma}

\begin{proof}
Let $c_k(p)=|\set{j\ge p:z_j\ge k}|$. By definition,
$r_p\ge k$ iff $c_k(p)\ge k$. Since $c_k(p)$ is non-increasing in $p$, the
qualifying positions form the prefix $0,\ldots,b_k$; hence $r_p\ge k$ iff
$b_k\ge p$. Qualifying thresholds also form a prefix, so their number is
exactly $r_p$. Feasibility for threshold $k+1$ implies feasibility for $k$,
which gives $b_{k+1}\le b_k$.
\end{proof}

Lemma~\ref{lem:breakpoint} changes the direction of the computation. Rather
than evaluating every suffix independently, we can test one threshold $k$
across many suffix positions simultaneously. Once $b_k$ is known, threshold
$k$ contributes to every position $p\le b_k$. This threshold-oriented view is
well suited to SIMT execution and underlies both GPU kernels below.

\begin{example}%
	\label{ex:suffix}
	For $z=(3,1,4,2)$, the only qualifying breakpoints are $b_1=3$
	and $b_2=2$. Lemma~\ref{lem:breakpoint} therefore gives the four
	suffix-H values $(2,2,2,1)$.
\end{example}

\subsubsection{Warp Kernel}
\label{sec:warp-kernel}

For segments of length $d\le\Warp$, one warp owns the complete segment.
Lane $p$ holds $z_p$. At threshold $k$, a warp ballot produces a bit mask
whose $j$th bit is set exactly when $j<d$ and $z_j\ge k$. Lane $p$ shifts
away positions $j<p$ and applies a population count, obtaining
$c_k(p)$, the number of qualifying values in its suffix. Thus all suffix
positions test the same threshold concurrently.

Let $M_k$ denote the ballot mask at threshold $k$. If
$\operatorname{popc}(M_k)<k$, then even the complete segment contains fewer
than $k$ values of size at least $k$. No suffix can therefore reach threshold
$k$, and no larger threshold can succeed. The warp can terminate uniformly at
this point. This tier uses only registers and warp-level primitives and
requires no block-level synchronization.

By Lemma~\ref{lem:breakpoint}, lane $p$ obtains $r_p$ by recording exactly the
thresholds whose shifted mask has population at least $k$. If
$h=r_0$, threshold $h+1$ fails even for the full segment, so the warp terminates
uniformly after all contributing thresholds have been processed.

\subsubsection{Tiled Block Kernel}
\label{sec:block-kernel}

A single warp cannot cover longer segments. Assigning one thread to each
suffix would duplicate most of the H-index work, while evaluating the
sequential recurrence with one thread would serialize the segment. We
therefore process longer segments with one block in tiles from right to left.

Suppose the suffix to the right of the current tile has already been
processed. The block carries three quantities for this suffix: a histogram
$\mathit{freq}[x]$ of its clamped values, its H-index $h_R$, and the count
$\cnt$ of values at least $h_R+1$. Let $h_L$ denote the H-index after the next
tile has been incorporated. Thresholds at most $h_R$ already qualify
throughout the new tile, so only thresholds in the band
$(h_R,h_L]$ can introduce new breakpoints.

For a threshold $k>h_R$, a parallel suffix scan within the tile computes, at
every tile position, the number of new values at least $k$. Adding the
corresponding count from the already processed suffix identifies the
breakpoint $b_k$. Between consecutive thresholds, the carried count for the
processed suffix is updated by
\begin{equation}
	\mathit{ge}[k+1]
	=
	\mathit{ge}[k]-\mathit{freq}[k].
	\label{eq:hist-update}
\end{equation}
Thus the histogram is populated during the right-to-left sweep but is never
rescanned in full for every threshold.

Across all tiles, each successful threshold is introduced only once. The
widths of the threshold bands therefore telescope to the final H-index of the
segment. This property is central to the work bound in
Section~\ref{sec:gpu-analysis}.

\begin{theorem}
	\label{thm:tiled}
	The right-to-left tiled block kernel computes
	\[
	r_p=\Hop(\mset{z_j:j\ge p})
	\]
	for every position of an arbitrary-length segment.
\end{theorem}

\begin{proof}
	Induct from right to left, carrying the processed suffix's histogram,
	H-index $h_R$, and count $\cnt$ above threshold $h_R+1$; the empty suffix
	initializes the invariant. For $k>h_R$, add the histogram-derived count
	$\mathit{ge}[k]$ to a parallel tile-suffix scan
	$|\set{j\in[p,e):z_j\ge k}|$. Their sum counts values at least $k$ in
	the complete suffix at $p$, so its rightmost qualifying position is
	exactly breakpoint $b_k$. Equation~\eqref{eq:hist-update} advances
	$\mathit{ge}[k]$ without rescanning the histogram. By
	Lemma~\ref{lem:breakpoint}, the first threshold failing throughout the
	tile ends its successful band; lower thresholds already qualify.
	The breakpoints determine every tile result, and folding the tile into
	the histogram restores the carried invariant.
\end{proof}

\subsection{Degree-Adaptive Kernels and Memory}
\label{sec:degree-tiers}

Temporal degree distributions can be highly skewed. Assigning every vertex to
the block kernel would underutilize the GPU for the many short segments, while
reserving enough shared memory for the largest segments would unnecessarily
reduce occupancy. We therefore partition vertices into static degree-tier
worklists during preprocessing.
Segments with $d_v\le32$ use the warp kernel. The tiled block kernel uses
$B=256$ threads: degrees $33$--$4096$ keep the histogram in shared memory,
and higher degrees use vertex-specific global scratch space. These
static worklists are constructed once and reused at every order. Degree
adaptation determines how a segment is processed, whereas canonical-state
compression determines which results persist to the next order.

\subsection{Correctness and Work--Span Analysis}
\label{sec:gpu-analysis}

Correctness follows by composition of the order-major recurrence with the warp
and tiled suffix-H kernels and Proposition~\ref{prop:compression}. The main
algorithmic trade-off is additional work to expose parallelism, which we
summarize in the work--span model.

\begin{theorem}
    \label{thm:work-span-segment}
    Let a block of $\Blk$ threads process a segment of length $d$, and let
    $h=r_0$. With the implemented iterative-doubling block scan, the tiled
    kernel has
    \[
    O\left((d+\Blk(h+1))\log\Blk\right)
    \]
    work,
    $
    O\left(\left(\left\lceil d/\Blk\right\rceil+h\right)\log\Blk\right)
    $
    span, and $O(d+\Blk)$ scratch storage. For fixed hardware block size, the
    work is $O(d+h)$ up to constant factors.
    These bounds apply to one segment processed by one block; they do not
    characterize scheduling or load balance across blocks or GPUs.
\end{theorem}

\begin{proof}
There are $\lceil d/\Blk\rceil$ tiles. Loading, folding, and histogram work is
$O(d)$. Each tile incurs one failed threshold test, while exactly $h$
successful threshold tests occur over the entire segment; each test performs a
scan over at most $\Blk$ entries with $O(\Blk\log\Blk)$ work and
$O(\log\Blk)$ span. Summing these terms gives the stated bounds.
\end{proof}

For the warp tier, at most $h+1\le\Warp+1$ threshold iterations are required;
with the hardware constant $\Warp=32$, this is bounded per segment. Thus the
GPU formulation may perform more work than the $O(nm)$ sequential sweep, but
replaces its loop-carried dependency with SIMT threshold tests and scans; the
overhead is governed by attained H-index values rather than segment degree
alone.

\section{Multi-GPU Execution}
\label{sec:multi-gpu}

The order-major formulation naturally supports an owner-computes
decomposition. Each vertex and its complete outgoing TCSR segment reside on
one GPU, so suffix-H computation is entirely local. Cross-device communication
is required only when a local temporal edge queries a previous-order canonical
state owned remotely. We therefore construct fixed, query-aware halos
containing only the distinct remote canonical states referenced by each GPU.
These routes are deduplicated during preprocessing and reused at every order.

\subsection{Ownership and Query-Aware Halos}
\label{sec:multi-ownership}

Let the $G$ GPUs be indexed by $0,\ldots,G-1$, and partition
\[
V=
V_0\mathbin{\dot\cup}\cdots\mathbin{\dot\cup}V_{G-1}.
\]
Write $\operatorname{own}(v)=g$ when $v\in V_g$. GPU $g$ owns all temporal
edges whose tail it owns,
$
\TE_g=
\set{
	e\in\TE:
	\operatorname{tail}(e)\in V_g
},$ with $
m_g=|\TE_g|.
$
Thus each vertex and its complete outgoing segment are colocated, and only the
gather phase can require remote state.
For distinct GPUs $g$ and $h$, define the receiver-specific route set
\begin{align}
	\mathcal R_{g\leftarrow h}
	=
	\{
	(v,p):&
	v\in V_h,\ p<d_v,\
	\exists e\in\TE_g
	\text{ with }\notag\\&
	\qquad\quad\operatorname{head}(e)=v,\
	\mathit{succ}[e]=p
	\}.
	\label{eq:route-set}
\end{align}
Each route represents one distinct canonical state requested by receiver $g$
from owner $h$. During preprocessing, every edge is mapped to an owner-local
state, the implicit zero sentinel, or a fixed incoming halo slot. Multiple
edges on the same receiver that query the same $(v,p)$ share one route and one
halo slot. Communication therefore depends on distinct cross-partition
canonical-state queries rather than directly on the number of cut temporal
edges.

\subsection{Execution and Correctness}
\label{sec:multi-execution}

Each GPU stores the two canonical-state buffers of its owned vertices and one
fixed slot for every incoming route. At order $i$, owner-local and sentinel
gathers execute immediately while owners concurrently push the requested
$F_{i-1}$ states into peer halo slots. After communication completes, the
remote-edge gathers consume these slots, followed by the degree-adaptive
suffix-H kernels on all owned segments. The resulting canonical states are
retained, the state buffers are swapped, and the next order begins. Since the
requested order $n$ is fixed, no convergence test or global reduction is
required between rounds.

Route destinations are fixed during preprocessing, so each outgoing route
stores the remote address of its halo slot. Owners can therefore write values
directly into peer memory without packed value buffers, transmitted slot
identifiers, or receiver-side scatter. We support \emph{full exchange}, which
writes every routed state at every order, and \emph{change-only exchange},
which writes a state only when its value differs from the value previously
sent on that route; otherwise the receiver retains its cached value.

Correctness follows directly from the single-GPU recurrence:
every gather reads either an owner-local state, the implicit zero sentinel, or
an exact halo copy, and complete outgoing segments remain on one GPU for the
subsequent suffix-H computation.

The full-pipeline implementation parallelizes preprocessing to avoid a serial host bottleneck. We reuse the device-side builder to construct the TCSR, successor indices, and canonical-state set using GPU sort/unique. For multi-GPU route construction, each receiver records the distinct remote canonical states it queries in a bit mask, avoiding sorting individual cut-edge witnesses. Under weighted-contiguous ownership, state owners are monotone in global-state order, so halo slots can be obtained by parallel rank queries on this mask. These structures are constructed once before the order rounds.\looseness=-1

\subsection{Communication and Memory}
\label{sec:multi-communication}

Let
$
R=
\sum_{g=0}^{G-1}
\sum_{h\ne g}
|\mathcal R_{g\leftarrow h}|
$
be the number of receiver-specific routes, and let
$
m_{\mathrm{cut}}
=
\left|
\set{
	e\in\TE:
	\operatorname{own}(\operatorname{tail}(e))
	\ne
	\operatorname{own}(\operatorname{head}(e))
}
\right|
$
be the number of cut temporal edges. A canonical state requested by two
different receivers contributes two routes, whereas arbitrarily many edges on
one receiver querying the same remote state contribute only one.

\begin{theorem}
	\label{thm:halo-communication}
	Full exchange sends exactly $R$ state values per order and $nR$ over $n$
	orders. Moreover,
		$R\le m_{\mathrm{cut}}$,
	and %
	the inequality can be strict by an arbitrarily large factor.
\end{theorem}

\begin{proof}
Each route is sent once per order and has at least one witnessing cut edge;
choosing one witness per route injects routes into cut edges. Many cut edges
may query the same remote %
state, making the inequality arbitrarily
strict.
\end{proof}

Thus $m_{\mathrm{cut}}$ is only an upper bound: deduplication may give
$R\ll m_{\mathrm{cut}}$. Change-only exchange further exploits
Lemma~\ref{lem:order-monotonicity}. A nonnegative canonical state
$q=(v,p)$ can strictly decrease at most $F_0[q]$ times, so one route transmits
at most
\begin{equation}
    N_q\le1+\min\{n-1,F_0[q]\}\le n.
    \label{eq:change-only-bound}
\end{equation}
Fixed halo slots eliminate transmitted state identifiers. Per-GPU storage is
linear in its owned edges and canonical states plus its incoming and outgoing
route metadata; the global temporal graph and canonical-state set are never
replicated on every GPU.

\subsection{Scaling and Ownership}
\label{sec:multi-runtime}
\label{sec:multi-partitioning}

Multi-GPU scaling is limited by two effects. First, a bulk-synchronous round is
limited by the most heavily loaded GPU. Because complete vertex segments are
owned by one device, a single high-degree vertex can dominate a partition;
splitting one vertex's suffix-H computation across GPUs would require a
different algorithm. Second, communication depends on query-aware boundary
structure: many cut edges may collapse to a single route, whereas the same
canonical state requested by several receivers must be delivered separately.
The relevant communication cost therefore depends on route counts, peer
connectivity, and transmitted bytes rather than on cut edges alone. Local
gathering is overlapped with halo exchange whenever possible.

The decomposition is valid for any ownership map. The weighted-contiguous
default uses $w_v=d_v(1+n)$; because $n$ is common to all vertices, this is
exactly contiguous edge balancing. A scan in vertex-ID order starts the next
GPU range when cumulative edge count reaches the next multiple of $m/G$.
Section~\ref{sec:evaluation} fixes this ownership to isolate exchange policy
from partitioning heuristics. Its monotone owners also enable rank/select
halo lookup.

\section{Experimental Evaluation}
\label{sec:evaluation}

We organize the evaluation around four questions:

\begin{itemize}
	\item \textbf{RQ1 Performance:} How do we compare with prior algorithms?
	
	\item \textbf{RQ2 Scalability:} How do we scale with input size, order, and hardware
	on controlled and real inputs?
	
	\item\textbf{RQ3 Ablation:} Which kernel, state, transition-time, and communication
	choices drive performance?
	
	\item\textbf{RQ4 Preprocessing:} What does preprocessing cost, and when is it
	amortized? 
 
\end{itemize}


\subsection{Experimental Setup}

All experiments ran on one node with two Intel Xeon Platinum 8481C processors
(104 physical cores) and eight NVIDIA H100 GPUs with 80\,GiB HBM3 each. Every
GPU pair was connected by NVLink, and peer access was enabled for multi-GPU
runs. We used CUDA 13.3, \texttt{nvcc} 13.3.73, and \texttt{g++} 11.4.0.
All C++17 methods use \texttt{g++ -O3} (and \texttt{-fopenmp} for the 104-core
engine); CUDA methods use \texttt{nvcc -O3}. CPU affinity uses
\texttt{OMP\_PLACES=cores} and \texttt{OMP\_PROC\_BIND=spread}; no explicit
NUMA policy is imposed. OM-CPU(104) uses parallel first-touch for round
buffers and working copies of lookup maps at every reported order; that copy
is included in preprocessing.

\paragraph{Methods}
The prior CPU algorithms are \textsc{Stream}, which computes all orders in one
reverse-time pass, and \textsc{Recurs}, the memoized definition-based method~\cite{oettershagen2023higherorder}.
Our implementations are sequential order-major (\emph{OM-CPU}), a 104-core
OpenMP implementation (\emph{OM-CPU(104)}), the single-GPU implementation
(\emph{OM-1GPU}), and the owner-computes implementation on eight GPUs
(\emph{OM-8GPU}). OM-CPU(104) uses one persistent thread team and static
edge-balanced vertex ranges. Unless stated otherwise, multi-GPU
runs use weighted-contiguous ownership, full exchange, and overlap.

\paragraph{Data}
The synthetic graphs are directed Erd\H{o}s--R\'enyi-style temporal
multigraphs with $|V|=m/25$. Each of the $m$ edges independently selects
a uniform source, a uniform distinct destination, and an integer timestamp
uniform in $[0,|V|/2]$; duplicate edges are retained and $\lambda_e=1$.
The uniform size sweep in Figure~\ref{fig:algorithm-scaling} spans
$10^5$--$10^9$ edges at $n=32$; eight-GPU measurements start at 50M.
Table~\ref{tab:large-multi-gpu} reports the matched 100M, 500M, and 1B
inputs for uniform and skewed sources. The order sweep fixes 10M edges
and extends to $n=1024$.
To isolate degree skew, a second controlled family instead samples source
rank by a seeded inverse-CDF power law with exponent $0.63$, while keeping
uniform distinct destinations, the same timestamp range, unit transitions,
and duplicate events. It has 100M, 500M, and 1B edges.
For real data, we use five normalized SNAP~\cite{snapnets}
networks and the Reddit reply network~\cite{Liu-2019-sampling};
Table~\ref{tab:real-datasets} gives their characteristics. We remove self-loops,
shift the first timestamp to zero, preserve direction and duplicate temporal
events, and use unit transition times except in the heterogeneous-$\lambda$
study. 
\begin{table}[t]
\caption{Real temporal-network characteristics. $Q/m$ is the canonical-state ratio for the outward index.}
\label{tab:real-datasets}
\centering\small
\begin{tabular}{lrrrrr}
\toprule
Dataset & $|V|$ & $m$ & $\Delta_{\max}$ & $Q/m$ & Ref.\\
\midrule
CollegeMsg & 1.9K & 59.8K & 1.1K & 0.502 & \cite{panzarasa2009patterns}\\
Email-Eu & 986 & 332.3K & 9.8K & 0.335 & \cite{paranjape2017motifs}\\
MathOverflow & 24.8K & 390.4K & 4.5K & 0.343 & \cite{paranjape2017motifs}\\
Reddit & 8.4M & 636.3M & 392.4K & 0.417 & \cite{Liu-2019-sampling}\\
StackOverflow & 2.58M & 47.9M & 71.7K & 0.368 & \cite{paranjape2017motifs}\\
WikiTalk & 1.09M & 6.1M & 233.5K & 0.412 & \cite{leskovec2010governance}\\
\bottomrule
\end{tabular}
\end{table}

\paragraph{Protocol and metrics}
Performance configurations have five fresh-process repetitions. Real-network
OM-CPU(104) configurations were randomly interleaved across datasets and
requested orders. We report medians
with deterministic nonparametric-bootstrap 95\%
confidence intervals; the deterministic convergence traces use one
instrumented process per graph. \emph{Algorithm time} is the time for the dependent
order rounds. \emph{End-to-end time} comprises parsing, preprocessing,
setup/upload, order rounds, and result extraction. 
We use a time-limit of one hour.
\begin{table*}[t]
\caption{Real-network order-round time (median processes) over five runs. \textsc{oot} denotes timeout. }\vspace{-2mm}
\label{tab:real-runtime}
\centering\scriptsize
\setlength{\tabcolsep}{3.2pt}
\renewcommand{\arraystretch}{0.94}
\begin{tabular*}{\textwidth}{@{\extracolsep{\fill}}lrrrrrrrr@{}}
\toprule
Dataset & $n$ & Unit & \textsc{Stream} & \textsc{Recurs} & OM-CPU & OM-CPU(104) & OM-1GPU & OM-8GPU \\
\midrule
CollegeMsg & 8 & ms & 77.90 & 489.1 & 4.037 & 4.868 & 1.398 & 3.413 \\
 & 32 & ms & 300.9 & 2,058.2 & 12.10 & 5.598 & 3.296 & 8.892 \\
 & 128 & ms & 1,327.8 & 10,779.5 & 40.51 & 9.632 & 8.681 & 28.49 \\
\addlinespace[1.2pt]
Email-Eu & 8 & ms & 2,546.6 & 26,348.6 & 16.17 & 0.937 & 12.86 & 13.15 \\
 & 32 & s & 12.08 & 103.5 & 0.0563 & 0.00260 & 0.0241 & 0.0268 \\
 & 128 & s & 67.79 & 553.5 & 0.2096 & 0.00889 & 0.0465 & 0.0652 \\
\addlinespace[1.2pt]
MathOverflow & 8 & ms & 1,376.3 & 22,756.3 & 24.43 & 1.201 & 5.904 & 7.077 \\
 & 32 & s & 5.872 & 94.97 & 0.0903 & 0.00327 & 0.0117 & 0.0176 \\
 & 128 & s & 30.74 & 504.1 & 0.3416 & 0.0106 & 0.0258 & 0.0411 \\
\addlinespace[1.2pt]
StackOverflow & 8 & s & 652.1 & \textsc{oot} & 4.401 & 0.1045 & 0.0914 & 0.0267 \\
 & 32 & s & 2,733.9 & \textsc{oot} & 17.10 & 0.3304 & 0.2698 & 0.0753 \\
 & 128 & s & \textsc{oot} & \textsc{oot} & 65.70 & 1.494 & 0.7737 & 0.2098 \\
\addlinespace[1.2pt]
WikiTalk & 8 & s & 45.82 & 772.9 & 0.3716 & 0.0187 & 0.0304 & 0.0233 \\
 & 32 & s & 211.1 & 3,425.1 & 1.400 & 0.0552 & 0.0962 & 0.0742 \\
 & 128 & s & 932.0 & \textsc{oot} & 5.351 & 0.2035 & 0.3248 & 0.2445 \\
\addlinespace[1.2pt]
Reddit & 8 & s & \textsc{oot} & \textsc{oot} & 92.65 & 1.985 & 1.894 & 0.353 \\
 & 32 & s & \textsc{oot} & \textsc{oot} & 352.3 & 8.765 & 5.537 & 1.085 \\
 & 128 & s & \textsc{oot} & \textsc{oot} & 1,326.0 & 27.906 & 17.491 & 3.556 \\
\bottomrule
\end{tabular*}
\end{table*}

\subsection{RQ1: Performance on Real Networks}
\label{sec:rq-performance}

OM-CPU consistently beats both prior methods: on StackOverflow at $n=32$,
it takes 17.10\,s versus 2733.9\,s for \textsc{Stream}
(Table~\ref{tab:real-runtime}). This algorithmic gain precedes GPU execution.
Parallel execution compounds this algorithmic gain on sufficiently large
graphs. On Reddit at $n=128$, eight GPUs achieve a
\RedditOrderOneTwentyEightEightVsCpuMt{} gain over OM-CPU(104). In
contrast, small graphs such as CollegeMsg do not contain enough work to
amortize multi-GPU synchronization, and one GPU is faster than eight.
The GPU does not always beat the multicore host: at $n\in\{32,128\}$ OM-CPU(104)
is faster than OM-1GPU on Email-Eu, MathOverflow, and WikiTalk. 
The distinction between round time and one-shot latency is important. On
Reddit at $n=128$, one- and eight-GPU runs using device-built canonical
states were interleaved without diagnostic kernel timing. The eight-GPU
end-to-end speedup over one GPU is \RedditOptimizedEteSpeedup{}
(bootstrap 95\% interval \RedditOptimizedEteCi{}). The confidence interval
for the process-wall speedup, \RedditProcessCi{}, includes parity.
The recurring order-round advantage is much larger than either one-shot gain.
Against OM-CPU(104), the eight-GPU Reddit end-to-end execution is
\RedditOptimizedVsCpuMtEte{} faster.

\subsection{RQ2: Scalability on Controlled Graphs}
\label{sec:rq-scalability}

\begin{figure}[t]
	\centering
	\includegraphics[width=\linewidth]{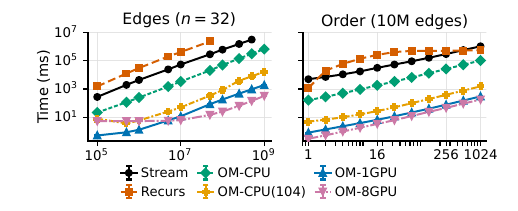}
	\caption{Controlled algorithm time. Left: edge count at $n=32$, with OM-8GPU
		measured from 50M. Right: order on 10M edges. Five-run medians and bootstrap
		95\% intervals are shown. \textsc{Recurs} times out at 100M and
		\textsc{Stream} at 1B; censored points are not extrapolated.}
	\label{fig:algorithm-scaling}
\end{figure}

Figure~\ref{fig:algorithm-scaling} isolates scaling with graph size and
requested order. Table~\ref{tab:large-multi-gpu} shows controlled
GPU strong scaling; real-network results are given below.

\begin{table*}[t]
\caption{Controlled order-32 scaling for uniform and skewed sources (five-run median ms). Multi-GPU runs use weighted-contiguous ownership and full overlapping exchange. $S_8$ compares the same multi-GPU engine on one and eight GPUs; CPU/8 compares OM-CPU(104) with eight GPUs. Halo is the ratio of median eight-GPU transfer time to median round time. %
}\vspace{-2mm}
\label{tab:large-multi-gpu}
\centering\scriptsize
\setlength{\tabcolsep}{5pt}
\begin{tabular}{lrrrrrrrr}
\toprule
Edges & OM-CPU(104) & 1 GPU & 2 GPUs & 4 GPUs & 8 GPUs & $S_8$ & CPU/8 & Halo \\
\midrule
\multicolumn{9}{l}{\textbf{(a) Uniform sources}} \\
100M & 836.1 & 188.1 & 84.6 & 42.1 & 25.0 & 7.52$\times$ & 33.4$\times$ & 33.7\% \\
500M & 7825.6 & 993.5 & 517.1 & 268.2 & 142.2 & 6.99$\times$ & 55.0$\times$ & 26.7\% \\
1B & 17145.4 & 2000.7 & 1081.1 & 573.1 & 322.0 & 6.21$\times$ & 53.2$\times$ & 26.8\% \\
\midrule
\multicolumn{9}{l}{\textbf{(b) Skewed sources}} \\
100M & 816 & 256 & 176 & 107 & 65.3 & 3.93$\times$ & 12.5$\times$ & 26.7\% \\
500M & 8749 & 1277 & 877 & 569 & 335 & 3.81$\times$ & 26.1$\times$ & 24.0\% \\
1B & 19398 & 2536 & 1746 & 1135 & 668 & 3.80$\times$ & 29.0$\times$ & 22.7\% \\
\bottomrule
\end{tabular}
\end{table*}

\textbf{Problem-size and order scaling.}
Algorithm time grows approximately linearly with both edge count and requested
order, as expected from the $O(nm)$ round complexity. The important difference
between methods is therefore the constant cost per edge-round. 
Against OM-CPU(104), OM-8GPU is 33.4--55.0$\times$ faster on the
100M, 500M, and 1B inputs (Table~\ref{tab:large-multi-gpu}). These are
low-skew, preprocessed order-round comparisons, not end-to-end speedups.
\begin{figure}[t]
\centering
\includegraphics[width=1\linewidth]{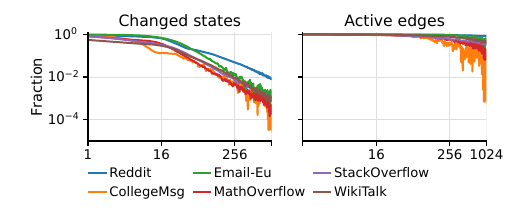}%
\caption{Convergence diagnostics on the six real graphs. Left: canonical
states changing at each completed order. Right: edges in segments with at
least one changed gathered input, hence eligible for recomputation at the
next order. None reaches a fixed point by order 1024.
}
\label{fig:convergence-real}%
\end{figure}
\begin{table}[t]
\caption{Reddit $n=128$ breakdown (median (s) over 5 runs).
GPU runs use device-built canonical states; OM-CPU(104) uses parallel
canonical construction, 104 successor workers, and parallel first-touch.
The parse-excluded row is the median of per-run total minus parse;
phase medians need not sum to the total.}
\label{tab:reddit-e2e-breakdown}
\centering\scriptsize
\setlength{\tabcolsep}{2pt}
\begin{tabular}{lrrrr}
\toprule
Phase & OM-CPU & OM-CPU(104) & OM-1GPU & OM-8GPU \\
\midrule
Parse & 40.82 & 41.55 & 33.78 & 34.03 \\
Sort/TCSR & 94.78 & 15.41 & 12.57 & 12.45 \\
Successor & 121.97 & 4.503 & 0.145 & 0.145 \\
State/worklist & 52.87 & 6.082 & 4.115 & 5.420 \\
Route & --- & --- & --- & 5.576 \\
Setup/H2D & --- & --- & 0.663 & 1.551 \\
Rounds & 1326.0 & 27.906 & 17.491 & 3.556 \\
D2H & 0.0281 & 0.0023 & 0.0027 & 0.0329 \\
\midrule
Total & 1636.5 & 95.88 & 69.50 & 64.86 \\
Without parse & 1596.0 & 54.36 & 35.72 & 31.10 \\
\midrule
Process wall & 1637.4 & 97.13 & 73.47 & 71.41 \\
\bottomrule
\end{tabular}%
\end{table}
Figure~\ref{fig:convergence-real} rules out a global fixed-point early exit
within the evaluated real-network orders. At order 128, the next round
would still process \FrontierRedditActiveAtOneTwentyEight{} of Reddit edges if whole segments formed a frontier.

\textbf{Hardware scaling.}
Additional hardware helps when each device receives enough local work.
OM-CPU(104) obtains \CpuHundredFourSpeedup{} speedup over one thread on the
10-million-edge case, while eight GPUs achieve
\LargeMultiSpeedupRange{} strong-scaling speedup on the 100M, 500M,
and 1B uniform inputs in Table~\ref{tab:large-multi-gpu}.
The maximum among these inputs, 7.52$\times$, occurs at 100M edges, whereas the 1B
eight-GPU speedup falls to 6.21$\times$
although its \LargeMultiBillionTransferShare{} halo-transfer share is lower
than at 100M.
Thus halo transfer alone does not explain the nonmonotone scaling; the
measurements do not isolate a single cause among workload balance, memory
behavior, and synchronization.

The real graphs expose a second limit that uniform controlled graphs do not:
segment skew. At order 32, OM-1GPU-to-OM-8GPU speedups are
5.10$\times$ on Reddit, 1.30$\times$ on WikiTalk, and 3.58$\times$ on
StackOverflow (Table~\ref{tab:real-runtime}). These compare the dedicated
single-GPU and multi-GPU implementations. Balancing edge counts is insufficient when
a very large vertex segment cannot be divided across devices: WikiTalk's
largest vertex contributes 30.7\% of one owner's edges.
The controlled skewed family makes that effect visible at larger scale
(Table~\ref{tab:large-multi-gpu}(b)). Eight-GPU within-engine speedup
is only 3.80--3.93$\times$, versus 6.21--7.52$\times$ for the uniform
family. At each matched size, the skewed inputs have lower halo-transfer
shares yet substantially weaker scaling, supporting the unsplittable-hub
explanation rather than an explanation based solely on transfer activity.
Construction dominates the larger skewed input, limiting its one-shot pipeline
advantage.

\subsection{RQ3: Ablation}
\label{sec:rq-ablation}

\textbf{Canonical-state compression.}
Compared with full edge-position state, canonical storage reduces working
memory by
\CompressionWorkingSavingRange{} and static memory by
\CompressionStaticSavingRange{} and improves round time by only
\CompressionComputeRange{}. On Reddit, the host-built full/canonical
comparison confounds storage and builder placement. 
In a
comparison using device-built layouts at $n=128$,
the ratio of full-state to canonical end-to-end time is
\RedditStorageFullOverCanonical{} (95\% interval
\RedditStorageFullOverCanonicalCi{}). Thus the full-state median is lower in this
comparison, but total latency does not distinguish the layouts: variation in
shared pipeline stages can exceed the layout difference.
Canonicalization's robust benefit is capacity, not a general round-time gain.

\textbf{Degree-adaptive kernels.}
Table~\ref{tab:review-ablation}(b) routes degree-$\leq32$ segments through
the block path. Order rounds slow by \KernelNoWarpSlowdownRange{}, reaching
1.893$\times$ on StackOverflow at $n=32$, where 90.4\% of nonempty segments
are short. Gather time remains nearly unchanged, while the slowdown occurs in suffix-H. Disabling warp early exit changes
total time by \KernelNoEarlySlowdownRange{} and has no material speedup.

\textbf{Static successors and breakpoint suffixes.}
Table~\ref{tab:review-ablation}(a) isolates two algorithmic choices without
changing the recurrence. Repeating binary successor searches every order
costs 3.29--12.42$\times$ on OM-CPU(104) across three real graphs and
1.03--1.78$\times$ on one GPU. The benefit is therefore especially large
on the multicore host. Independent per-suffix GPU evaluation is not
uniformly inferior on tiny inputs: it can be faster at order 8 on
CollegeMsg and MathOverflow. 

\begin{table}[t]
\caption{Order-round ablations on real graphs. (a) Algorithmic choices; 
	Values above one indicate a slowdown. Per-round successor search repeats temporal binary searches, independent suffix computes each suffix separately, and hub-grid makes grid-wide threshold passes over the large WikiTalk segment. (b) block-only/default GPU ratios when short segments ($d_v\leq32$) bypass the warp tier. Short shares refer to nonempty vertices and edges at $n=32$.}
\label{tab:review-ablation}
\centering\scriptsize
\setlength{\tabcolsep}{3pt}
\textbf{(a) Algorithmic choices}\par\smallskip
\begin{tabular}{llrrr}
\toprule
Ablation & Graph & $n$ & OM-CPU(104) & OM-1GPU \\
\midrule
Per-round successor & MathOverflow & 32 & 3.29$\times$ & 1.03$\times$ \\
 & WikiTalk & 32 & 3.94$\times$ & 1.08$\times$ \\
 & StackOverflow & 32 & 12.42$\times$ & 1.78$\times$ \\
\addlinespace[1pt]
Independent suffix & CollegeMsg & 8 & --- & 0.597$\times$ \\
 & CollegeMsg & 32 & --- & 0.962$\times$ \\
 & MathOverflow & 8 & --- & 0.857$\times$ \\
 & MathOverflow & 32 & --- & 1.70$\times$ \\
\addlinespace[1pt]
Hub-grid & WikiTalk & 32 & --- & 1.76$\times$ \\
 & WikiTalk & 128 & --- & 1.31$\times$ \\
\bottomrule
\end{tabular}
\par\medskip
\textbf{(b) Warp tier}\par\smallskip
\begin{tabular}{lrrrr}
\toprule
Dataset & Short $V$\% & Short $E$\% & $n=8$ & $n=32$ \\
\midrule
Email-Eu & 26.6 & 0.7 & 1.009$\times$ & 1.019$\times$ \\
MathOverflow & 91.1 & 14.9 & 1.074$\times$ & 1.140$\times$ \\
Reddit & 76.2 & 4.9 & 1.112$\times$ & 1.161$\times$ \\
StackOverflow & 90.4 & 18.0 & 1.674$\times$ & 1.893$\times$ \\
WikiTalk & 89.9 & 9.0 & 1.131$\times$ & 1.165$\times$ \\
\bottomrule
\end{tabular}
\end{table}

\textbf{Query-aware exchange.}
Deduplication reduces distinct remote queries to approximately 0.857 per cut
edge on uniform controlled graphs and 0.682 on Reddit. At eight GPUs,
change-only exchange reduces transmitted values by \ExchangeSavedBillion{}
on the billion-edge graph at order 32 and \ExchangeSavedReddit{} on Reddit
at order 128, but the corresponding full/change-only round-time ratios are
only \ExchangeTimeBillion{} and \ExchangeTimeReddit{}. These
workload-specific results show that substantial traffic savings need not yield
comparable runtime gains.
At 50M edges, no-overlap has the lowest median under either policy, but
confidence intervals overlap. In a separate comparison of ownership policies
on Reddit at $n=128$, round-robin lowers eight-GPU rounds from 3.551 to 2.951\,s,
yet increases the full pipeline from 65.0 to 79.7\,s because route
construction becomes more expensive. Weighted-contiguous ownership therefore
remains the one-shot headline configuration, even though round-robin has
the better rounds. Rank/select halos exploit monotone owners.

\textbf{Nonuniform transition times.}
Across six heterogeneous-$\lambda$ variants on MathOverflow, WikiTalk,
and StackOverflow, round-time ratios are \TransitionRoundRatioRange{} of
unit-transition runs: $\lambda_e$ changes static successors, not the
order-round computation.

\subsection{RQ4: Preprocessing and Amortization}
\label{sec:rq-preprocessing}

On Reddit at $n=128$, rounds account for only 5.5\% of eight-GPU
end-to-end time (Table~\ref{tab:reddit-e2e-breakdown}). Input processing
and static construction therefore dominate one-shot latency, despite
device-built TCSR and canonical states, sort-free halo planning, and
parallel host construction with first-touch.

Our end-to-end measurements include parsing a normalized, preindexed text
edge list, excluding download and normalization. This is the largest individual
eight-GPU phase on Reddit. Its cost depends on the application: data may already
be available in a binary representation close to the required layout, reducing
decoding and conversion work. The reported times characterize our text-input
pipeline.

At $n=32$, the eight-GPU pipeline is at best at parity with one GPU on
MathOverflow, WikiTalk, and StackOverflow; its measured one-shot gain
is specific to high-order Reddit. Reuse requires the \emph{same} outward
graph and $\lambda$: order $n$ already yields lower orders, changed
$\lambda$ requires new successors, and inward indexing requires a
transpose rebuild. Reported times assume no reuse.

\subsection{Limitations}
Measurements are limited to one NVLink-connected node. Small workloads and unsplittable hub segments limit overhead amortization and load balance, weakening scaling under degree skew. Input processing and static construction also reduce one-shot gains. Hardware performance counters were unavailable, preventing direct attribution of scaling to DRAM bandwidth or memory efficiency.

\section{Conclusion}

The order-major reformulation improves sequential complexity, exposes parallelism, and identifies sufficient cross-order state. Experiments show substantial acceleration at scale, while preprocessing limits one-shot gains and small workloads and degree skew limit GPU scaling. 
Future work will investigate cooperative multi-block processing of high-degree vertex segments to reduce hub-induced load imbalance without the repeated scans and synchronization overhead observed in our grid-wide variant.

\section*{Acknowledgment}
This research was supported by the NVIDIA Academic Grant Program using NVIDIA H100 GPUs and the NVIDIA CUDA Toolkit.

\balance

\FloatBarrier


\FloatBarrier

\end{document}